\documentclass[prb,groupedaddress,twocolumn]{revtex4}
\usepackage{graphicx}

\begin{document}
\title{
Quasi-linear magnetoresistance in an almost two-dimensional band structure
}
\author{A. J. Schofield}
\affiliation{School of Physics and Astronomy, University of
Birmingham, Edgbaston, Birmingham, B15 2TT, United Kingdom}
\author{J. R. Cooper}
\affiliation{IRC in Superconductivity, Madingley Road, Cambridge, CB3 0HE,
United Kingdom}
\date{15 February 2000}
\begin{abstract}

We present a theoretical study of the orbital magnetoresistance in a
unixial anisotropic metal within the relaxation-time
approximation. The appearance of a new dimensionless scale,
$\delta=4t_\perp/\epsilon_F$, allows the possibility of a new region
at intermediate fields where the magnetoresistance is linear in
applied magnetic field for currents flowing along the unixial
direction.  (Here, $t_\perp$ characterizes the bandwidth along the
unixial direction.) In the limit of large anisotropy (small $\delta$),
corresponding to a quasi-two-dimensional metal made up of weakly coupled
layers, we obtain an analytic expression for the magnetoresistance
valid for all magnetic fields. We test our analytic results numerically and 
we compare our expressions with the $c$-axis
magnetoresistance of $\rm Sr_2RuO_4$.

\end{abstract}
\maketitle

\section{Introduction}
A growing number of compounds have been synthesized whose crystal
structure consists of weakly coupled metallic layers. Foremost among
these are the cuprate metals where the two-dimensional nature
electronic structure has provoked much speculation about the nature of
the resulting metallic state. The issue of whether in-plane
excitations can move coherently between copper oxide planes remains
contentious~\cite{strong}. Several other metallic compounds with a
layered structure---for example organic compounds based on the
bis-ethylenedithiotetrathiafulvalene molecule and other metallic
oxides such as $\rm Sr_2RuO_4$---do possess a Fermi surface which is
shaped like a slightly warped cylinder~\cite{mackenzie}.  In this
paper we study magnetotransport in a quasi-two-dimensional (2D) metal
in order to provide a benchmark against which more exotic types of
behavior can be compared.  Specifically we study out-of-plane
transport using the relaxation-time approximation in the presence of
an in-plane magnetic field. Our main result is that there is a
``Kapitza'' region~\cite{kapitza} where the transverse
magnetoresistance $\Delta \rho_c/\rho_c$ is proportional to the
applied magnetic field.  We justify this with an analytic expression
for the magnetoresistance in the quasi-two-dimensional limit which is
valid for {\em arbitrary} magnetic fields.  We also obtain an
expression for the magnetoresistance with any degree of unixial
anisotropy that can be evaluated numerically.

The study of magnetotransport within the Boltzmann formalism is long
established and is well described in a number of classic
texts\cite{abrikosov,pippard}. Interpreting magnetoresistance
measurements is complicated by the fact that the magnetoresistance is
identically zero for an isotropic metal. The amount of anisotropy
determines the measured magnetoresistance, which is therefore rather
sensitive to the detailed properties of the material. Analytic results
are usually limited to the very weak or very high-magnetic-field
regimes. At low fields the Zener-Jones expansion yields a
magnetoresistance quadratic in the magnetic field with a coefficient
depending on the variation of the mean-free-path around the Fermi
surface.  At high fields the magnetoresistance saturates when current
flow is along closed Fermi-surface directions or maintains a quadratic
field dependence for currents along open Fermi-surface directions (see
page 118 of Ref.~\onlinecite{abrikosov}). To our knowledge, there have
been no analytic expressions for the magnetoresistance of a realistic
bandstructure which interpolate between these known limits.

In this paper we present a calculation which, while respecting the
high- and low-field results mentioned above, also applies at
intermediate fields where we find a linear magnetoresistance. We have
obtained an analytic expression for the magnetoresistance, valid in
the limit of strong anisotropy, which we believe to be the first
straightforward example of a magnetoresistance formula valid for all magnetic
fields. This result should prove useful in characterizing the
properties of quasi-two-dimensional metals using 
transport measurements. The paper is organized as
follows. We first present a simple calculation of the
magnetoconductance that illustrates how having a warped cylindrical
Fermi surface can give rise to a linear magnetoresistance. We then
present a more formal solution of the Boltzmann equation and derive
the conductivity tensor for arbitrary levels of anisotropy. Our
analytic result emerges as a limiting case. Finally we discuss the
implications of these results for experiment and compare with known
data.

\section{Simple picture}

We consider a metal with the following dispersion relation
\begin{equation}
\epsilon(\vec k)= {\hbar^2 \over 2m_\circ}\left(k_x^2 + k_y^2 \right)
- 2t_\perp
\cos \left(k_z c \right) + 2t_\perp\; .
\label{dispersion}
\end{equation}
We have adopted the customary notation that directions in reciprocal
space are labeled $x$, $y$, and $z$ while the corresponding 
real space lattice is defined by the $a$, $b$ and $c$ directions. 
It describes free particles in the $ab$ plane coupled by a
perpendicular transfer integral, $t_\perp$ in the $c$ direction to
adjacent planes. The magnitude of $c$ gives the spacing between 
planes that can be
combined with $t_\perp$ to form an effective mass for out-of-plane
motion 
\begin{equation}
m_\perp = \hbar^2/2 t_\perp c^2 \; .
\label{defmperp}
\end{equation} 
(This is the $z$-axis band mass in the limit $\epsilon_F \ll t_\perp$
when the Fermi surface forms a closed spheroid.)

\begin{figure}
\includegraphics[width=\columnwidth]{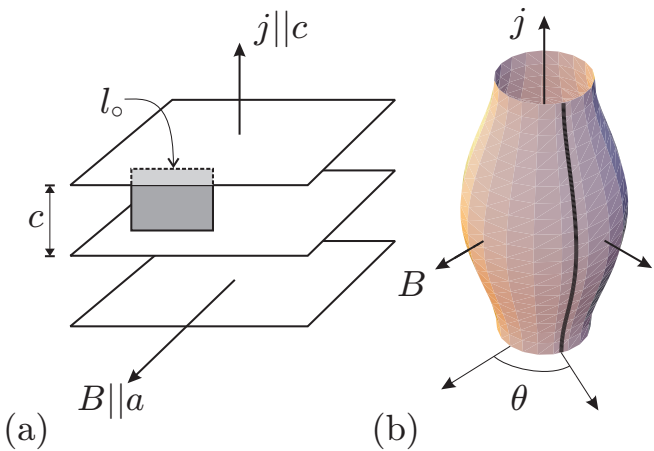}
\caption{The field geometry and approximate quasiparticle
orbits---here shown for $\delta=2/3$.  
(a) The magnetic field lies in
the plane and the current flows out of the plane. The linear
magnetoresistance regime occurs when the magnetic flux through a loop
defined by the in-plane mean-free-path, $l_\circ$, and the plane
separation, $c$, becomes of the order of a flux quantum or greater.
(b) In the limit of large anisotropy, we can approximate the
quasiparticle orbits as lines on the Fermi surface with constant
radial angle $\theta$ and constant in-plane velocity $v_\circ$.  This
gives an approximate derivation of the low-field ($B^2$) and
intermediate-field ($|B|$) regions.}
\label{approxorbits}
\end{figure}

We begin with an approximate derivation of the new regime by
considering the magnetoconductance to lowest order in 
$t_\perp$.  Chambers' expression~\cite{chambers} for the components of
the conductivity tensor in a magnetic field within the relaxation-time
approximation is 
\begin{equation}
\sigma_{ij} = {e^2 \over 4 \pi^3} \oint {dS \over \hbar |\vec{v}|}
\int_0^{\infty} v_i(0) v_j(t) e^{-t/\tau} dt \; .
\label{chambers}
\end{equation}
For each area element of the Fermi surface, $dS$, we integrate  
the velocity $\vec{v}(t)$ measured along a semiclassical quasiparticle
orbit. These orbits are defined by the Lorentz equation of
motion
\begin{equation}
\hbar {d\vec{k} \over dt} = -e \vec{v} \times \vec{B} \; ,
\label{motion}
\end{equation}
where $\vec{v}=\vec{\nabla}_k \epsilon(\vec{k})/\hbar$. 
In this paper we will be interested in configurations where the
magnetic field is parallel to the planes and the current is flowing
perpendicular to the planes [see Fig.~(\ref{approxorbits}a)].

To lowest order in $t_\perp$
we can ignore both the $z$-axis dispersion in the equation of motion
and the closed orbits. The rate of change of ${\vec k}$ then
only depends on the angle $\theta$ between the magnetic field and the
in-plane Fermi velocity of an electron [see Fig.~(\ref{approxorbits}b)].  
So we have
\begin{equation}
\hbar {d k_z \over dt} = -e v_\circ B \sin \theta \; ,
\end{equation}
where the in-plane Fermi velocity, $v_\circ$, is constant here.
Hence we see that
\begin{equation}
c k_z(t) = c k_z(0) + \Omega_c t \sin \theta \; .
\end{equation}
We can determine how the velocity in
the $c$ direction changes as the quasiparticle moves across the Fermi
surface by combining the above result with Eq.~(\ref{dispersion}) giving
\begin{equation}
v_c(t) = {2t_\perp c \over \hbar} 
\sin\left[k_z c  + \Omega_c t \sin\theta \right] \; .
\end{equation}
Here we have defined a ``cyclotron'' frequency
\begin{equation}
\Omega_c = {e v_\circ B c \over \hbar} \;,
\end{equation}
which is the fastest rate at which 
quasiparticles traverse the Brillouin zone.
It turns out that this sets the scale of the crossover from weak
($\Omega_c \tau \ll 1$) to intermediate fields ($\Omega_c \tau \gtrsim 1$).

Substituting this velocity, $v_c(t)$, into the expression for the
conductivity Eq.~(\ref{chambers}), we may integrate over time and $k_z$
to give the out-of-plane conductivity expressed as an integral over
the orbits, namely,
\begin{equation}
\sigma_c (B) = {e^2 t_\perp^2 \tau k_F c \over \pi^2 \hbar^3 v_\circ}
\int_0^{2\pi} {d \theta \over 1 + \Omega_c \tau^2 } \; .
\end{equation}
Integrating this is straightforward and yields the following 
magnetoconductance
\begin{equation}
{\Delta \sigma_c(B) \over \sigma_c(0)}
 = {1 \over \sqrt{1 + \Omega^2_c \tau^2}} \; .
\label{simpleapprox}
\end{equation}

At low fields ($\Omega_c \tau \ll 1$) this gives the usual quadratic
field dependence as obtained by the Zener-Jones expansion.  However in
the intermediate field regime $(\Omega_c \tau \gtrsim 1)$ the
magnetoconductivity falls off as $1/B$. This is the essential result
of this paper. It arises because there is a range of cyclotron
frequencies for traversing the Brillouin zone. 
Note too that the magnetoconductance becomes
universal, independent of the degree of anisotropy and depending only
on the in-plane properties.  We can emphasize this by writing the
cross-over condition in terms of the magnetic length $l_B^2 =
\hbar/eB$, namely $\Omega_c \tau=1$ corresponds to 
\begin{equation}
{l_\circ c \over l_B^2} = 1 \; .
\end{equation} 
So the linear region is reached when approximately one flux quantum threads
an area formed by the in-plane mean-free-path ($l_\circ = v_\circ \tau$)
and the interplanar spacing.  For a typical layered oxide ($c \sim
12$\AA) one can expect to see this regime at 10 Tesla when the
in-plane mean free path reaches around 500 \AA. As discussed later,
experiments~\cite{hussey_1997} on $\rm Sr_2RuO_4$
provide evidence for the validity of this
expression. 

However, this cannot be the complete story since very general
arguments show that $\sigma_{c}$ must go as $1/B^2$ at high
fields\cite{abrikosov}. Since the conductivity is the sum of the
conductivities from all orbits, the high-field form cannot be
recovered simply by including the contribution of the closed orbits: a
$1/B^2$ contribution from closed orbits will never dominate the $1/B$
from open orbits. The correct high field result emerges when we
consider higher-order effects in $t_\perp$ for both the open and
closed orbits in the exact solution of section~III. We will see
that the high field $1/B^2$ regime occurs when $\Omega_c \tau \gg
\sqrt{\epsilon_F/t_\perp}$.

A further reason for a more detailed treatment is that the
magnetoresistance is identically zero for an ellipsoidal Fermi surface
with a constant $\tau$. This is because of a cancellation
between the Hall and magnetoconductance. We therefore would like to
verify that there is no such cancellation here.  We will do this
through an exact solution of the Boltzmann equation in which we
compute all components of the conductivity tensor and hence the
magnetoresistance.

\section{Solving the Boltzmann Equation}

\begin{figure}
\includegraphics[width=\columnwidth]{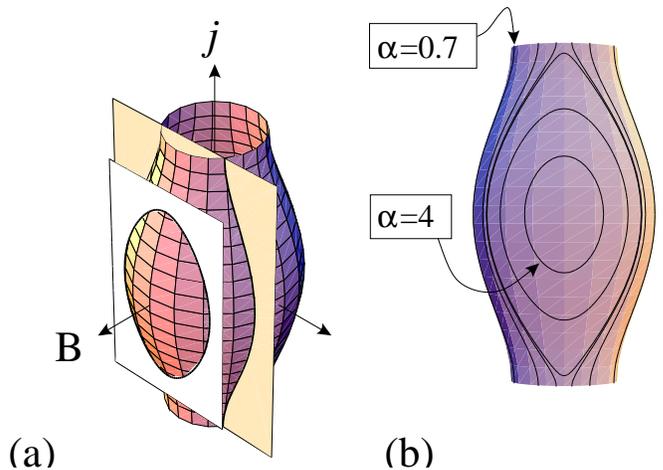}
\caption{The exact orbits over the Fermi surface with $\delta=2/3$.
(a) The quasiparticle moves along lines defined by the
intersection of the Fermi surface and planes perpendicular to the
field direction.  These orbits can be either open or closed and are
described by Jacobi elliptic functions.
(b) We can label these orbits by a
parameter $\alpha$, the parameter of the Jacobi elliptic
function. Here we illustrate orbits where 
$\alpha=0.7, 0.9, 0.99, 1.01, 1.5$ and $4$. 
For $\delta \le \alpha \le 1$ the orbits are
open. For $1 \le \alpha < \infty$ the orbits are closed.}
\label{exactorbits}
\end{figure}

Our treatment will follow closely that of Abrikosov~\cite{abrikosov}.
In the previous section we only treated the quasiparticle orbits
approximately.  Since the Lorentz force in Eq.~(\ref{motion}) acts
perpendicular to the electron motion, energy is conserved and the
electron is constrained to move along a constant energy line with fixed
momentum in the $\vec{B}$ direction. Here we take $B$ to be parallel
to $a$.  For finite $t_\perp$, $\theta$ is no longer constant along
the orbits and there are some closed orbits as illustrated in
Fig.~(\ref{exactorbits}a).

Quasiparticles at the Fermi surface determine the transport properties and
we may identify two regions depending on the degree of anisotropy. We
introduce a parameter which is a measure of this anisotropy
\begin{equation}
\delta = {4 t_\perp \over \epsilon_F} \; .
\label{defdelta}
\end{equation}
If $\delta > 1$, the Fermi surface is closed and the trajectory of all
quasiparticles in momentum (and real) space follows closed loops.
While the system may have an anisotropic effective mass, the
qualitative features of transport will not be much modified from a
typical three-dimensional metal. If $\delta < 1$ then we still have
some closed orbits but there are now some trajectories that extend
across the Brillouin zone in the $c$ direction [see
Fig.~(\ref{exactorbits}a)].  Only these orbits have been treated in
section II (and then only approximately). We now aim for a more
complete analysis.

To consider the conductivity tensor we must solve the Boltzmann
equation.  Rather than use the momentum components $k_x$, $k_y$ and
$k_z$, in the presence of a magnetic field it is more convenient to
work in terms of a new coordinate system $k_x, \epsilon,$ and
$t$. Here $t$ is the time taken to move along the momentum orbits
defined by the equation of motion, Eq.~(\ref{motion}).  The advantage
of this coordinate system is that the magnetic field is included
implicitly and does not appear in the Boltzmann equation.  Within the
relaxation-time approximation, one may write the Boltzmann equation
as~\cite{abrikosov},
\begin{equation}
{\partial \psi \over \partial t} + {\psi \over \tau} = e\vec{E} 
\cdot \vec{v} \; ,
\label{boltz}
\end{equation}
where the electron distribution function has been written as $f =
f^{(0)}-\psi \partial_\epsilon f^{(0)}$.  This first-order
differential equation may be solved straightforwardly. 

Substituting into Eq.~(\ref{motion}) gives
\begin{eqnarray} 
dt &=& - {m_\circ \hbar \over eB} {dk_z \over 
\sqrt{2m_\circ \epsilon - \hbar^2 k_x^2 + 4m_\circ t_\perp 
\left[\cos 
\left({k_z c} \right) -1 \right]}} \; ,\\
&=& -{\sqrt{\alpha} \over \omega_c}{c \over 2 } 
{dk_z \over \sqrt{1- \alpha \sin^2\left({k_z c /2}\right)}}
\; , \label{integrable}
\end{eqnarray}
where
\begin{equation}
\alpha = {8 t_\perp m_\circ \over 2 m_\circ \epsilon_F - \hbar^2 k_x^2}
\; .
\label{defk}
\end{equation}
We have defined a new ``cyclotron'' frequency~\cite{cyclotron}
\begin{equation}
\omega_c = {e B c \over \hbar} \sqrt{2 t_\perp \over m_\circ} 
= {eB \over \sqrt{m_\circ m_\perp}} = \Omega_c \sqrt{\delta \over 4} 
\;,
\end{equation}  
the fastest rate at which quasiparticles perform closed orbits.  This
is the natural scale for cyclotron motion perpendicular to the plane. As
might be anticipated from a Bohr quantization picture, it also sets the
scale for Landau-level quantization and hence the quantum effects
which signal the breakdown of quasi-classical transport theory. 
The variable $\alpha$ labels each cyclotron orbit. It is bounded from
below by $\delta$
and we will use it to substitute for $p_a$
\begin{equation}
\hbar k_x = \sqrt{2 m_{\circ} \epsilon - 8 t_\perp m_{\circ}/\alpha} \; .
\end{equation}  
The orbits are open for $\delta < \alpha < 1$ and closed for $ 1 <
\alpha < \infty$. 

It is very unusual that Eq.~(\ref{integrable}) is both integrable
{\em and} its solution is invertible so that a closed form expression
for the orbits may be found. The solution can be written in terms of
the Jacobian elliptic functions
\begin{eqnarray}
k_z(t) &=& -{2 \over c} {\rm JacobiAmplitude}\left[{
\omega_c t \over \sqrt{\alpha}}, \alpha \right],\\
v_c(t) &=& -{2 \hbar \over c m_\perp} {\rm JacobiCN}\left[{
\omega_c t \over \sqrt{\alpha}}, \alpha \right] {\rm JacobiSN}\left[{
\omega_c t \over \sqrt{\alpha}}, \alpha \right], \nonumber \\
&& \\
v_b(t) &=& {2 \hbar \over c \sqrt{\alpha m_\perp m_{\circ}}}
{\rm JacobiDN}\left[{
\omega_c t \over \sqrt{\alpha}}, \alpha \right]. 
\end{eqnarray} 
These equations exactly describe the quasiparticle's motion over the
Fermi surface defined by Eq.~(\ref{dispersion}) in the presence of a
magnetic field along the $a$ direction. We have adopted the notation
of Mathematica~\cite{wolfram} and Abromowitz and Stegun~\cite{AS} in
using the parameter $\alpha$ rather than the modulus, $k=\sqrt{\alpha}$,
to define these functions.

These periodic functions play the role of the trigonometric functions
that appear in the solution of the spherical problem. To make this
more explicit, we can map the elliptic functions when the parameter
$\alpha$ is greater than one (describing closed orbits) to those with a
parameter $\beta = 1/\alpha$ less than one~\cite{AS}. 
We may then write
\begin{eqnarray}
v_c(t) &=& -{2 \sqrt{\beta} \hbar \over c m_\perp}
{\rm JacobiDN}[\omega_c t, \beta]
{\rm JacobiSN}[\omega_c t, \beta] \; , \\
&& \lim_{\beta \rightarrow 0} -{2\hbar \over c} {\sqrt{\beta} \over
m_\perp} \sin (\omega_c t) \; , \\
v_b(t) &=& {2 \hbar \over c} \sqrt{\beta \over m_\perp m_{||}}
{\rm JacobiCN}[\omega_c t, \beta] \; , \\
&=& \lim_{\beta \rightarrow 0} {2\hbar \over c} \sqrt{\beta \over
m_\perp m_\circ} \cos (\omega_c t) \; , 
\end{eqnarray}
The limiting case of $\beta \rightarrow 0$ describes all of the orbits
when the Fermi surface becomes spheroidal ($\delta \rightarrow
\infty$). 

To compute the conductivities we use the solution of the Boltzmann
equation [Eq.~(\ref{boltz})] and compute the current. This gives the
Chambers' formula [Eq.~(\ref{chambers})] which may be written
as (see Ref.~[\onlinecite{abrikosov}])
\begin{equation}
j_\alpha = {2e^2B \over \left(2\pi \hbar\right)^3} 
\int_{-p_a^0}^{p_a^0} \! \! \! \! \! \! dp_a 
\int_0^{T(p_a)} \! \! \! \! \! \! \! \! dt 
\int_{-\infty}^t \! \! \! \! \! \! \!  dt' 
v_\alpha(t) v_\beta(t')
E_\beta e^{(t'-t)/\tau} \; .
\label{currents}
\end{equation}
This triple integral can be drastically simplified when we recall that
the orbits are all periodic and so have a well-defined Fourier
series. The Fourier series for the Jacobian elliptic functions are all
tabulated~\cite{GR}. This allows one to do the integral over $t'$ and,
using the orthogonality of the components of the Fourier series, one
can also do the integral over $t$. The algebra is somewhat tedious but
the result is that the conductivity can be expressed as a rapidly
convergent sum followed by a single integral over $p_a$.  

We will use the definition of the tensor conductivity 
$j_{\alpha} = \sigma_{\alpha\beta}E_\beta$ with the magnetic field
along the $a$ direction. The only 
nonzero components of the conductivity tensor will be $\sigma_{aa}$,
$\sigma_{bb}$, $\sigma_{cc}$ and $\sigma_{bc}=-\sigma_{cb}$. 
There is no longitudinal magnetoresistance for the dispersion
of Eq.~(\ref{dispersion}) so $\sigma_{aa}$ is unaffected
by the magnetic field and, by rotational symmetry in the $ab$ plane,
will be equal to $\sigma_{bb}$ in the absence of a magnetic field. We
can simplify some of the expressions by introducing a number of
parameters: a universal conductivity and the effective-mass ratio
\begin{equation}
\sigma_0 = {2 e^2 \tau \over c^3 m_\perp} \; , \quad \quad 
r = {m_\circ \over m_\perp} \; .
\end{equation} 

\begin{widetext}

For the exact solution of the conductivity tensor we find 
the following components
\begin{eqnarray}
\sigma_{aa} &=& {\sigma_0 \over \pi^3 \sqrt{\delta}} \left[
4 \int_{\min(1,\delta)}^1 
{K(\alpha) \sqrt{\alpha - \delta} \over \alpha^2}
d\alpha  + 
\int_0^{\min(1,\delta^{-1})} 
K(\beta) \sqrt{1 - \beta \delta}  d\beta \right] \;,  
\label{saa} \\ 
\sigma_{bb}&=&{\sigma_0 \sqrt{\delta} \over \pi} \left[ 
\int_{\min(1,\delta)}^1 {1 + 8 S_1 (\alpha, \omega_c
\tau / \sqrt{\alpha})
\over K(\alpha) \sqrt{\alpha - \delta}} d\alpha + 
8 \int_0^{\min(1,\delta^{-1})} 
{\tilde{S}_1 \left(\beta,\omega_c \tau \right) \over 
K(\beta) \sqrt{1-\beta\delta}} d\beta \right] \; , 
\label{sbb} \\
\sigma_{cc}&=&{2 \pi r\sigma_0\sqrt{\delta} } \left[ 
4 \int_{\min(1,\delta)}^1 {S_2 (\alpha, \omega_c
\tau / \sqrt{\alpha})
\over K(\alpha)^3 \alpha^3\sqrt{\alpha - \delta}} d\alpha + 
\int_0^{\min(1,\delta^{-1})} 
{\tilde{S}_2 \left(\beta,\omega_c \tau \right) \over 
K(\beta)^3 \sqrt{1-\beta\delta}} d\beta \right] \; .
\label{scc} 
\end{eqnarray}
\end{widetext}
The remaining Hall component is given simply by
\begin{equation}
\sigma_{bc} = {eB \tau \over m_\circ} \sigma_{cc} \; .
\label{sbc}
\end{equation}
In these expressions, $K(\alpha)$ is the 
elliptic integral and we have defined the following
sums that involve, $q$,  the nome of the elliptic integral
$q(\alpha) = \exp[-\pi K(1-\alpha)/K(\alpha)]$ 
\begin{eqnarray}
S_1(\alpha, x) &=& \sum_n \left[ {q(\alpha)^n \over 1 +
q(\alpha)^{2n}} \right]^2 {1
\over 1 + \left( {n \pi x \over K(\alpha)} \right)^2} \; , 
\label{s1} \\
S_2(\alpha, x) &=& \sum_n \left[{n q(\alpha)^n \over 1 +
q(\alpha)^{2n}} \right]^2 {1
\over 1 + \left( {n \pi x \over K(\alpha)} \right)^2} \; .
\label{s2}
\end{eqnarray}
The summations above are over the positive integers ($n=1,2, \cdots
\infty$).  The sums $\tilde{S}_1$ and $\tilde{S}_2$ are the same as
$S_1$ and $S_2$ respectively except they are summed over the
positive half-integers ($n = 1/2, 3/2, 5/2, \cdots, \infty$).

Equations~(\ref{saa})~to~(\ref{s2}) are the exact solution for electrical
transport within the relaxation-time approximation and are valid for
arbitrary values of $\epsilon_F$ and $t_\perp$. 
To make further progress and to make contact with the result of the
simple calculation outlined previously, we need to work in the limit
of large anisotropy: $\delta \ll 1$. In this limit, the dominant term in
the conductivity tensor comes from $\sigma_{cc}$ [Eq.~(\ref{scc})] and,
in particular the small $\alpha$ range of the integral. 
It is therefore a good approximation to replace 
the integrand with its small $\alpha$ limit [i.e. take only
the first term in the sum, and let $K(\alpha) \rightarrow \pi/2$]. 
Doing this gives the following approximate form for the
conductance
\begin{equation}
\sigma_{cc}(B) \sim {\sigma_0 r \over 2 \pi^2}
\arctan\left({\sqrt{1/\delta - 1} \over \sqrt{1 + (\Omega_c
\tau)^2}} \right) {1 \over \sqrt{1 + (\Omega_c \tau)^2}} \; .
\label{badapprox}
\end{equation}
This result recovers the form of our approximate derivation of the
result [Eq.~(\ref{simpleapprox})] but remains correct in the extreme
high field
limit. With $1/\sqrt{\delta} \gg \Omega_c \tau \gg 1$ we can replace
$\arctan$ by $\pi/ 2$ and we have the linear magnetoresistance
regime as before. However, the correct $1/B^2$ asymptote is recovered
when $\Omega_c \tau \gg 1/\sqrt{\delta}$.   

\begin{figure}
\includegraphics[width=\columnwidth]{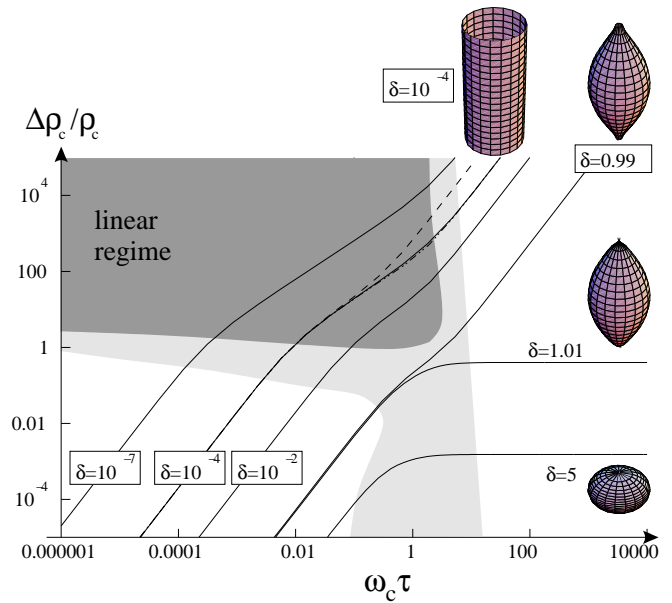}
\caption{A numerical treatment of the magnetoresistance for arbitrary
fields and Fermi surface topologies. These are computed using the
results of Eqs.~(\ref{saa}) to~(\ref{s2}). In the limit of $\delta \gg 1$
the magnetoresistance vanishes as expected in limit of a spheroidal
Fermi surface.  The previously known asymptotic regimes are unshaded
[always $B^2$ at weak fields with a high-field asymptote that is
$B^2$ for open Fermi surfaces ($\delta <1$) or saturates for a closed
Fermi surface ($\delta >1$).] The linear regime is heavily shaded
while the crossovers are lightly shaded.  The dashed
line shows the result (for $\delta = 10^{-4}$) from the lowest Fourier
component of the open orbits [Eq.~(\ref{badapprox})]. Adding a
numerical factor based on the asymptotic limit from all orbits
[Eq.~(\ref{goodapprox})] is also shown for this $\delta$ but is
indistinguishable from the numerical result. Note that $\delta <
10^{-2}$ for all three bands of ${\rm Sr_2RuO_4}$.}
\label{numerics}
\end{figure}

This is a better approximation than our previous treatment because we
are giving an exact treatment of the lowest Fourier component of the open
orbits. We are dividing the conductivity into a sum over quasiparticle
orbits and now each of these becomes a Fourier series. The additive
nature of the conductivity and the Fourier series means that each
component must contain the physics of the high-field asymptotic limit
as well as the linear intermediate field regime. We can compare the
role of higher-order Fourier components which are neglected in
deriving Eq.~(\ref{badapprox}) by comparing with numerical treatment of
equations~\ref{saa}~to~\ref{s2}. The dashed line in
Fig.~\ref{numerics} shows the magnetoresistance keeping only the
lowest order component [Eq.~(\ref{badapprox})] for $\delta=10^{-4}$. The
deviation from the numerically exact result indicates where high order
components become important. However since all of these higher order
terms contribute to the $B^2$ asymptote, they can be taken into
account by including an extra numerical factor in Eq.~(\ref{badapprox}) so
that the numerical and analytic results match in the high field, small
$\delta$ limit. This gives the following interpolating expression
\begin{equation}
{\Delta \rho_c \over \rho_c} \simeq 
{\pi \sqrt{1  + (\Omega_c \tau)^2 } \over 
2\arctan (1/\sqrt{\delta  + 0.0263\,\delta (\Omega_c \tau )^2})
}  - 1 \; ,
\label{goodapprox}
\end{equation}
where the factor of $0.0263$ is obtained by summing all orbits
numerically in the high field and small $\delta$ limit.  This function
is plotted in Fig.~\ref{numerics} but is virtually indistinguishable
from the numerical result. The appearance of a new region where the 
magnetoresistance is linear in $B$ 
may clearly be seen (Fig.~\ref{numerics}) as the dispersion
becomes more two dimensional.

For completeness we consider also the Hall resistivity with current
along $c$ and the Hall voltage being developed along the $b$ direction.
We find
\begin{equation}
\rho_{H} = {- \sigma_{bc} \over \sigma_{bb} \sigma_{cc} + \sigma_{bc}^2}
\simeq -{e B \tau \over m_\circ \sigma_{bb}} \; .
\end{equation}
Thus there is no anomalous regime in the Hall resistivity, which
remains linear at all magnetic fields and reflects the carrier
concentration in the usual way. 

\section{Comparison with experiment}

\begin{figure}
\includegraphics[width=\columnwidth]{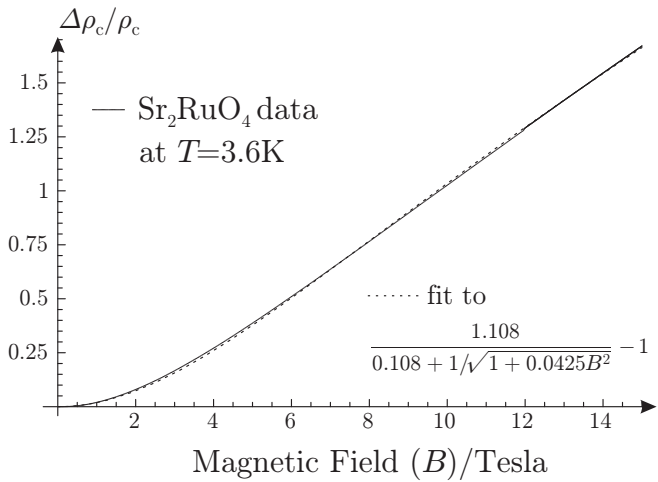}
\caption[]{The transverse $c$-axis magnetoresistance of ${\rm
Sr_2RuO_4}$ (data after Hussey {\it et al.},
Ref.~\onlinecite{hussey_1997}).  This clearly shows the linear regime
we have predicted. The development of the regime at relatively low
values of $\Delta \rho_c/\rho_c \sim 0.2$ as opposed to $\sim 1$ is an
indication that more than one band with differing mean-free-paths are
involved in the $c$-axis conductivity. This is demonstrated by the
fit, which includes a band with low c-axis conductivity and short
mean-free-path. This is consistent with the band structure
(Ref.~\onlinecite{bergemann_1999}) and the Hall coefficient
(Ref.~\onlinecite{hallsr214}).}
\label{sr2ruo4}
\end{figure}

${\rm Sr_2RuO_4}$ is probably the best characterized two-dimensional
metal. The current experimental $c$-axis
magnetoresistance~\cite{hussey_1997} clearly shows the linear
magnetoresistance regime we have discussed [see Fig.~(\ref{sr2ruo4})].
We now discuss what quantitative information we may determine from
this.

Detailed de Haas van Alphen studies give a very clear picture of the
degree of warping of the Fermi-surface
sheets~\cite{bergemann_1999}. These are expressed as variations in the
radius of the Fermi surface in the $xy$ plane as the Brillouin zone
is traversed along the $z$ direction. To translate to our notation we
note that
\begin{equation}
{\Delta k_F \over k_F} \simeq {2 t_\perp \over \epsilon_F} = {\delta \over
2} \; .
\end{equation}
In general, terms in the $c$-axis dispersion can involve $\cos (\nu k_z
c)$ where $\nu$ is an integer.  Thus far we have considered $\nu =
1$. Terms with $\nu >1$ increase the $c$-axis conductivity which
depends on the square of the $c$-axis velocity ($\sim
\nu^2$). Furthermore, there can also be an angular variation of the
$c$-axis dispersion, $t_\perp$, within the plane. This not only
modifies the numerical prefactor in the conductivity but generally
means that the $c$-axis magnetoresistance becomes dependent on the
orientation of the field within the $ab$ plane.  This is known to be
important in the cuprate superconductors~\cite{xiang_1996}.  Finally,
for the particular case of ${\rm Sr_2RuO_4}$, there are three bands
which give additive contributions to the total conductivity.

Because of the three bands and angular variation of $t_\perp$, 
it is misleading to use the
formulas we have developed to determine a quantitative measure of 
Fermi-surface anisotropy from
the magnetoresistance. Indeed we have argued that the
magnetoresistance becomes universal for each band in the $\delta
\rightarrow 0$ limit. Instead we can use the experiments as a probe
of in-plane properties---the mean-free-path. 

To obtain a good fit to data while introducing a minimum of free
parameters, we consider 
\begin{equation}
\sigma_{cc} (B) = \beta_1 + {\beta_2 \over \sqrt{1 + \alpha^2_1 B^2}} \; .
\end{equation}
This represents one fluid of electrons ($\beta_1$) with a short
mean-free-path, $l \ll l_B^2/c$, and therefore insensitive to the
fields, and second ($\beta_2$) with a much longer mean-free-path
[following Eq.~(\ref{simpleapprox})]. The magnetoresistance depends
only on $\beta_1/\beta_2 \sim 0.1$ and $\alpha_1 \sim
\sqrt{0.04}$. This is consistent with de Haas van Alphen
measurements~\cite{bergemann_1999} which suggest that one Fermi
surface sheet, $\gamma$, is considerably less dispersive in the
$c$-direction than the other two.  In addition, the assumption of a
small mean-free-path on that sheet is also consistent with the Hall
coefficient that remains strongly temperature dependent in the regime
of this experiment~\cite{hallsr214}.  $\alpha_1$ may be related to the
mean-free-path using $l_\circ = \alpha_1 \hbar /ec$ which gives a
value of $l_\circ \sim 2000$\AA~at 3K on the two sheets with the most
z-axis dispersion.  This is consistent with the observation of
unconventional superconductivity in this sample at around
$1K$~\cite{scsr214}.

\section{Conclusion}
So to summarize: we have given an exact solution for electrical
transport within a quasi 2D band structure. In doing so we find that
the new dimensionless parameter $\delta = 4 t_\perp/\epsilon_F$ is
important.  This leads to a new region ($\sqrt{\delta} \lesssim
\omega_c \tau \lesssim 1$) in the magnetoresistance where the $c$-axis
transverse magnetoresistance is large $(\Delta \rho_c/\rho_c
\gtrsim1)$ and linear in the applied field.  An asymptotically exact
expression for the magnetoresistance has been obtained in the limit of
small $\delta$, i.e. the limit of weakly coupled 2D planes. This new
region has been observed at low temperatures in the quasi 2D metal
$\rm Sr_2RuO_4$. For the over-doped thallium cuprate there are signs
that one is beyond the low field regime at 11 Tesla~\cite{hussey2}.

\section{Acknowledgements}
Some of the work described in this paper was done in
collaboration with J. M. Wheatley. 
We have also benefitted from numerous discussions with N. E. Hussey,
D. E. Khmelnitskii, A. P. Mackenzie and A. J. Millis.  On completion of
this work we became aware of a related paper by Lebed and
Bagmet~\cite{lebed}. This work complements their approach by computing
an analytic expression for the magnetoresistance and including the
effect of closed orbits.


\end{document}